\documentclass[5p]{elsarticle}
\usepackage[breaklinks]{hyperref}
\usepackage[utf8]{inputenc} 
\usepackage{lipsum} 
\usepackage{rotating} 
\usepackage{color}	
\usepackage{listings} 
\usepackage{enumitem} 
\usepackage[breakwords]{truncate}
\usepackage{subcaption}
\usepackage[export]{adjustbox} 
\usepackage{tabularx}
\usepackage{booktabs}
\usepackage{multirow}
\usepackage[normalem]{ulem}
\usepackage{algorithm}
\usepackage{algpseudocode}
\usepackage[pdftex,dvipsnames]{xcolor}

\useunder{\uline}{\ul}{}

\hypersetup{
	bookmarks=true,         
	pdffitwindow=true,     
	pdfstartview={FitH},    
	pdftitle={Learning from the Past: a Process Recommendation System for Video Game Projects using Postmortems Experiences},    
	pdfauthor={Cristiano Politowski, Lisandra M. Fontoura, Fabio Petrillo, Yann-Gaël Guéhéneuc},     
	colorlinks=true,       
	linkcolor=red,          
	citecolor=green,
	filecolor=magenta,      
	urlcolor=cyan           
}

\definecolor{mygray}{rgb}{0.5,0.5,0.5}
\lstdefinestyle{code}{ 
	basicstyle=\scriptsize\ttfamily,
	frame=lines,
	breaklines=true,
	tabsize=2,
	numberstyle=\tiny\color{mygray},
	numbers=left,
	float 
}
\lstdefinestyle{md}{ 
	basicstyle=\scriptsize\ttfamily,
	frame=lines,
	breaklines=true,
	tabsize=2,
	float 
}

\newcommand{\projpage}[1]{\href{https://polako.github.io/rec-sys-proc-games/}{https://polako.github.io/rec-sys-proc-games/}}

\makeatletter

\journal{Information and Software Technology}

\begin{document}

\begin{frontmatter}

\title{Learning from the Past: a Process Recommendation System for Video Game Projects using Postmortems Experiences}

\tnotetext[mytitlenote]{All the material of this project, including the extracted processes, video game contexts, and validations data is available on-line at \projpage{}.}

\author[mymainaddress]{Cristiano Politowski\corref{mycorrespondingauthor}}
\cortext[mycorrespondingauthor]{Corresponding author}
\ead{cpolitowski@inf.ufsm.br}

\author[mymainaddress]{Lisandra M.\ Fontoura}

\author[petrilloadr]{Fabio Petrillo}

\author[petrilloadr]{Yann-Gaël Guéhéneuc}

\address[mymainaddress]{Departamento de Computação Aplicada (DCOM), Universidade Federal de Santa Maria, Santa Maria, RS, Brasil \\ E-mail: \{cpolitowski,lisandra\}@inf.ufsm.br}

\address[petrilloadr]{Department of Computer Science \& Software Engineering, Concordia University, Montréal Quebec H3G 1M8, Canada \\ E-mail: fabio@petrillo.com, yann-gael.gueheneuc@concordia.ca}

\begin{abstract}
\noindent\textit{Context:} The video game industry is a billion dollar industry that faces problems in the way games are developed. One method to address these problems is using developer aid tools, such as Recommendation Systems. These tools assist developers by generating recommendations to help them perform their tasks.

\noindent\textit{Objective:} This article describes a systematic approach to recommend development processes for video game projects, using postmortem knowledge extraction and a model of the context of the new project, in which ``postmortems'' are articles written by video game developers at the end of projects, summarizing the experience of their game development team. This approach aims to provide reflections about development processes used in the game industry as well as guidance to developers to choose the most adequate process according to the contexts they're in.

\noindent\textit{Method:} Our approach is divided in three separate phases: in the the first phase, we manually extracted the processes from the postmortems analysis; in the second one, we created a video game context and algorithm rules for recommendation; and finally in the third phase, we evaluated the recommended processes by using quantitative and qualitative metrics, game developers feedback, and a case study by interviewing a video game development team.

\noindent\textit{Contributions:} This article brings three main contributions.
The first describes a database of developers' experiences extracted from postmortems in the form of development processes. The second defines the main attributes that a video game project contain, which it uses to define the contexts of the project. The third describes and evaluates a recommendation system for video game projects, which uses the contexts of the projects to identify similar projects and suggest a set of activities in the form of a process.

\end{abstract}


\end{frontmatter}

\section{Introduction}

The video game industry is a multi-billionaire market, which revenues have been growing up through the years. According to the digital marketing company Newzoo, in 2016, the video game industry achieved a revenue of US\$99.6 billions, 8.5\% more than in 2015, expecting US\$118 billions in 2019 \cite{Newzoo:2016}.

Video game development is a highly complex activity \cite{game_plan}. Studies about the video game industry concluded that video game projects do not suffer from technical problems but face management and processes problems \cite{Petrillo:2008,Petrillo:2009}.

Problems like ``unrealistic scope'', ``feature creep''\footnote{\textit{Feature creep} occurs in any software project when new functionalities are added during the development phase, thus increasing the project size \cite{Petrillo:2009}} and ``cut of features'' \cite{Petrillo:2008,Petrillo:2009} are recurrent in game development. In addition, because of lack of maturity, development teams often do not adopt systematic processes \cite{Murphy-Hill:2014} or they use traditional approaches, such as the waterfall process \cite{Politowski:2016}.
Other problems were identified in the IGDA annual report \cite{Weststar:2016} in which, for example, 52\% of the interviewees answered ``yes'' when asked if ``crunch time''\footnotemark~is necessary during game development.

\footnotetext{\textit{Crunch time} is the term used in the video game industry for periods of extreme work overload, typically in the last weeks before the validation phase and in the weeks that precede the final deadline for project delivery \cite{Petrillo:2009}.}

To help game developers minimize these problems, researchers proposed approaches to apply agile practices in game development \cite{Kanode:2009,Petrillo:2010,Keith:2010}. However, recent work showed that at least 35\% of the  development teams continue using hybrid processes (mixing waterfall and agile) or \textit{ad-hoc} processes \cite{Politowski:2016}.
Other researchers have tried, through a survey, understand the problems faced by game developers and the role of software engineering in the game industry \cite{politowskisoftware}.
Their results showed that
(1) the use of systematic processes, opposed to the development lacking any process, had a correlation to project success and
(2) the most frequent problems were \textit{delays}, \textit{non realistic scope}, and \textit{lack of documentation}.
They also analyzed 20 postmortems and extracted their corresponding software processes and showed that the adoption of agile processes is increasing, corresponding to 65\% of the identified processes \cite{Politowski:2016}.

Consequently, we claim that game development teams must improve their project management practices by adopting systematic processes. We also claim that an approach to improve processes is to learn from past project problems. Thus, we use as main source of information ``postmortems'', which are articles written by game developers at the end of their projects, summarizing the experiences of the game development team \cite{goodbye_postmortem}. However, to be a useful source of information for developers, the non-structured data contained in the postmortems must be structured. The developers should also be warned of possible problems that may occur during development, given the lack of video game developers' knowledge about software process \cite{Petrillo:2009}.
Thus, we want to develop a recommendation system to support video game development teams in building their development processes, recommending a set of activities and characteristics from past projects with similar contexts.

\begin{figure}[ht]
	\centering
	\includegraphics[width=1\linewidth]{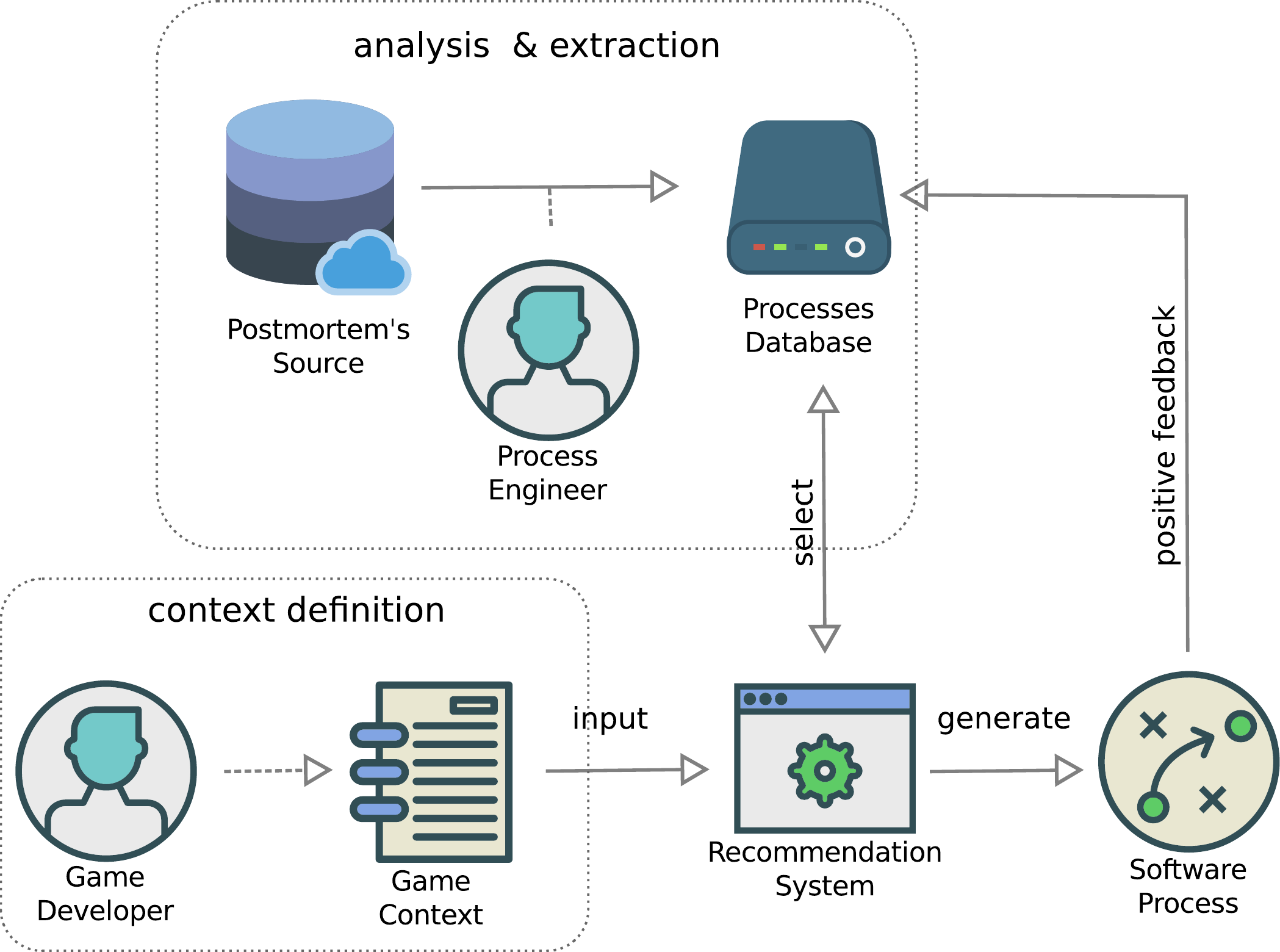}
	\caption{Summary of the approach.}
	\label{fig:contrib}
\end{figure}

We describe an approach to build a recommendation system to recommend \textbf{software processes}, using \textbf{context} and a \textbf{similarity degree} with previous game development projects. We want the recommendation system to support game developers in choosing a sequence of activities and practices that fit within their contexts. Figure \ref{fig:contrib} summarizes our approach: (1) the construction of the \textbf{processes database} with characteristics extracted from postmortems; (2) the definition of the \textbf{video game context} to characterize video game projects; and (3) the building of a \textbf{recommendation system} that suggests the \textbf{software process} using the defined context. Finally, we evaluate the recommendation system quantitatively, qualitatively and by performing a case study with a video game development team.

The remainder of this paper is structured as follows.
Section \ref{sec:background} describes the background of our approach.
Section \ref{sec:rmethod} describes the method used to build and validate the recommendation system.
Section \ref{sec:dev} explains how we build and apply the recommendation system.
Section \ref{sec:validation} describes the validation of the recommended processes.
Section \ref{sec:discussion} discusses our approach, the recommendation system, and its recommendations.
Section \ref{sec:threats} summarizes threats to the validity of our validation.
Section \ref{sec:related} describes the related work.
Section \ref{sec:conclusion} summarizes our approach, its results, our contributions, and future work.

\section{Background}
\label{sec:background}

This section briefly describes the concepts necessary to understand the rest of the article.

\subsection{Recommendation Systems}
\label{sec:b-rs}

Recommendation Systems (RS) are programs that use data to build models to help users in making decision. They recommend items based of explicit or implicit users' preferences. They help in alleviating the problem of information overload, showing users the most interesting items according to some criteria, offering relevance and diversity \cite{robillard2010recommendation}.

For example, e-commerce Web sites use recommendation systems to suggest their customers items.
They may also be used in technical domains as well to help developers in their tasks.
They are then labeled Recommendations Systems for Software Engineering (RSSE).
They offer informations regarding software development activities.
RS are divided in four different groups \cite{robillard2014recommendation}:

\begin{enumerate}
\item \textbf{Collaborative filtering}: A technique for generating recommendations in which the similarity of opinions of agents on a set of items is used to predict their similarity of opinions on other items.

\item \textbf{Content-based}: A technique for creating recommendations based on the contents of the items and the profiles of the users' interests.

\item \textbf{Knowledge-based}: A technique for providing recommendations using knowledge models about the users and the items to reason about which items meet users' requirements.

\item \textbf{Hybrid}: A technique to generate recommendations that combines two or more of the previous techniques to build its recommendations.
\end{enumerate}


RSSE have been used to a wide range of purposes. For example,
\textit{source code within a project}, helping developers to navigate through project's source code;
\textit{reusable source code}, providing new discoveries to programmers, like classes and functions;
\textit{code examples}, elucidating coders who does not know how implement a particular library / framework;
\textit{issue reports}, gathering and providing knowledge about past issues;
recommending \textit{tools, commands and operations} to solve certain problems in software projects with big scope;
and recommending the best \textit{person} to a perform a task.

Our problem is similar to these previous scenarios, particularly with \textit{issue reports, recommending operations and persons}, because we want to help developers by recommending processes based on previous successful and unsuccessful video game projects, described in form of postmortems. Thus, we will use the \textit{collaborative filtering} technique, however, instead of \textit{opinions of agents} we will use game project details to compare the similarities between video game projects.

\subsection{Software Process}
\label{sec:b-sp}

Humphrey \cite{Humphrey:1988} defined a software process as a set of activities, methods, and practices used in the production and evolution of software systems.
There are many process types available in the literature, both academic and professional. Fowler \cite{fowler2001new} divides software processes into \textbf{predictive} and \textbf{adaptive} processes, according to their characteristics. Predictive processes emphasize sequential flows of activities and requirements definitions before the beginning of software development. Activities are defined beforehand. Such processes require strong requirements definitions.
\textit{Waterfall} is an example of this type of processes \cite{Royce:1970}. Adaptive processes imply short development cycles performed continuously, delivering ready-to-use features at the end of each cycle \cite{Larman:2003}. They emphasize continuous improvement of the processes \cite{fowler2001new}. Examples of this type of processes are \textit{Scrum} \cite{schwaber2002agile} and \textit{Extreme Programming} \cite{beck2000extreme}.

Some authors claim that adaptive or agile development has been used since the beginnings of software development  \cite{larman2003iterative}, meanwhile others see the adoption of agile methods as a ``natural evolution'' in the way software is developed \cite{fowler2001new}. In spite of that, the fast adoption of agile processes is notorious, principally after the Agile Manifesto in 2001\footnotemark. According to the annual report report made by \textit{Version One} \cite{VersionOne2015}, in 2016, 96\% of respondents claimed making use of agile practices. A fact also evidenced in the video games area, where the agile methodologies are majority \cite{Politowski:2016, Musil:2010}.

\footnotetext{\href{http://agilemanifesto.org}{http://agilemanifesto.org}}

\subsection{Software Context}
\label{sec:b-sc}

Knowing a project context---its characteristics---is crucial to the success of a game development project. However, the variety of situations in which software projects are used is expanding, making existing premises outdated \cite{fuggetta2014software}.

However, to the best of our knowledge, there has been only few attempts to categorize projects contexts.
Boehm \cite{Boehm:2003} defined five characteristics that distinguish predictive processes from agile processes: size, criticality, personnel, dynamism and culture.
Kruchten \cite{kruchten2013contextualizing} defined an approach to contextualize software projects by defining characteristics of agile projects in a model called \textit{Octopus}. This model defines eight characteristics for software projects: size, criticality, business model, architecture, team distribution, governance, rate of change and age of the system.

Both previous works tried to contextualize software projects. However, video game projects, although software as well, have greater multidisciplinarity \cite{Petrillo:2008, Petrillo:2009} and, consequently, we claim that these small sets of characteristics do not properly define video game software development projects.

\subsection{Video Games Postmortems}
\label{sec:b-pm}

Video games postmortems are documents that summarize the developers' experiences of a game development project after (either successful or unsuccessful) completion. They emphasize positive and negative outcomes of the development \cite{goodbye_postmortem}. They are usually written right after the ends of game development projects by managers or senior developers \cite{Callele:2005}. They are an important tool for knowledge management and sharing \cite{Fridley2013}. They help teams to learn from their own experiences and plan future projects and also learn from the successes and problems of other teams. They are pieces of knowledge that can be reused by any development team and they capture examples and real-life development experiences. They are so important that some authors argue that no project should finish without postmortem \cite{Fridley2013}.

Usually, postmortems follow the structure proposed in the \textit{Open Letter Template} \cite{pma} and divide in three sections. The first section summarizes the project and presents some important outcomes of the project. The next two sections discuss the two most interesting characteristics of any of game development:

\begin{itemize}
\item \textbf{What went right} discusses the {\em best practices} adopted by developers, solutions, improvements, and project management decisions that help the project. All these characteristics are critical in planning future projects.

\item \textbf{What went wrong} discusses difficulties, pitfalls, and mistakes experienced by the development team in the project, both technical and managerial.
\end{itemize}

It is important to note that postmortems are not exclusively about process. As a matter of fact, this kind article is, most of the time, very informal and does not contain any useful information regarding the development process. Given its free structure, authors used to write about a wide range of subjects including game design, game balancing, \textit{gameplay} and many others details that do not add to development process details.

\subsection{Principal Component Analysis}
\label{sec:b-pca}

Principal Component Analysis (PCA) reduces the dimensionality of a set of data constituted of many related variables, keeping the majority of the variance contained in the data \cite{jolliffe2002principal}.

\begin{figure}[ht]
    \includegraphics[width=1\linewidth]{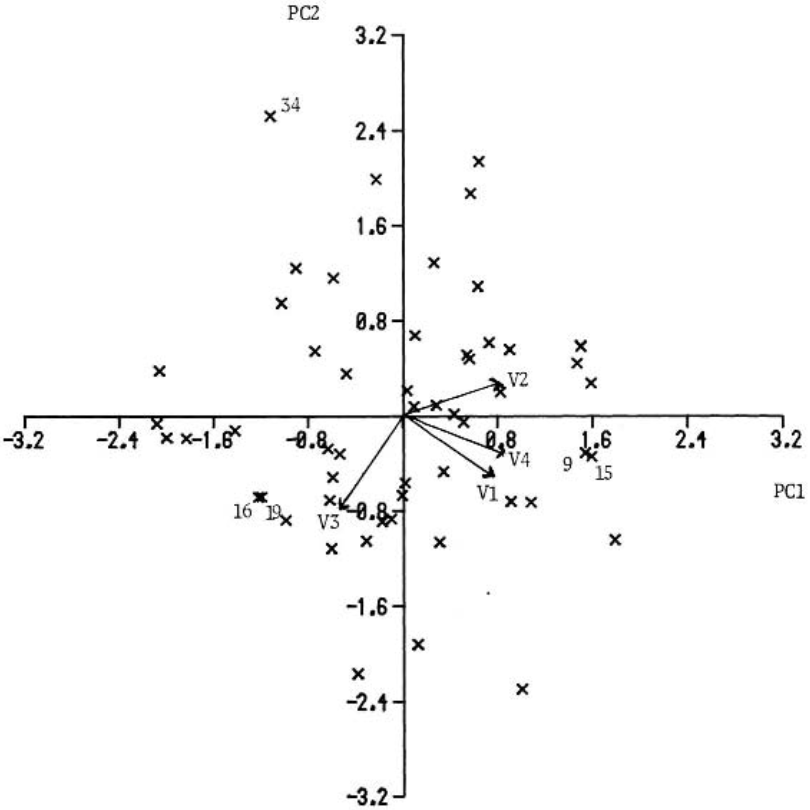}
    \caption{PCA biplot example.}
    \label{fig:biplot_example}
\end{figure}

PCA is useful when there is a large set of variables because the data can be plotted in two dimensions, allowing a straightforward visual representation, instead of comparing numbers. A plot of the first two sets of variables, that is, a \textit{biplot}, provides the best possible data representation in two dimensions, useful to find patterns as well \cite{jolliffe2002principal}. Figure~\ref{fig:biplot_example} shows an example of \textit{biplot} with four variables: V1 to V4. It is important to note that the more close a sample is from another, the more correlated they are, that is, more similar.

\section{Research Method}
\label{sec:rmethod}

Our approach, illustrated by Figure \ref{fig:contrib}, is inspired by the ``design concerns'' related to build recommendation systems described by Robillard \textit{et al.} \cite{robillard2014recommendation}. It consists in four challenges:

\begin{enumerate}
\item \textit{Data preprocessing}. Transform raw data in a sufficiently interpreted format. For instance, source code has to be parsed or commits have to be aggregated. See Section~\ref{sec:m1}.

\item \textit{Capturing context}. Different from traditional RS which are dependents of user profiles, RSSE focus on \textit{tasks}. In this regard, a \textit{task context} is all information which the recommender has access to produce recommendations. This information is generally incomplete and/or noisy. It is described in Section~\ref{sec:m2}.

\item \textit{Producing recommendations}. It is the algorithms' work on the data. Different strategies can be used, however, the most suitable approach for RSSE is \textit{Collaborative Filtering}. See Section~\ref{sec:m3} for details.

\item \textit{Presenting the recommendations}. Presents the items of interest, that is, \textit{activities} in our case. Described in Section~\ref{sec:m4}.
\end{enumerate}

\subsection{Data and Preprocessing}
\label{sec:m1}

Because of constraints, in particular trade secret, to access real documentation and artifacts from video game development projects, we gathered data from public-domain source: postmortems from game development Web sites, like \textit{Gamasutra}\footnote{https://www.gamasutra.com/}, and \textit{Wikipedia} entries of game projects. We observed that only \textit{Gamasutra} publishes postmortems about video games. Its dataset is large enough to fulfill our needs in this article. Moreover, all postmortems offered by \textit{Gamasutra} follow the same format, thus a single source of data helps mitigating the lack of formal structure of postmortems.

\begin{figure}[ht]
    \includegraphics[width=1\linewidth]{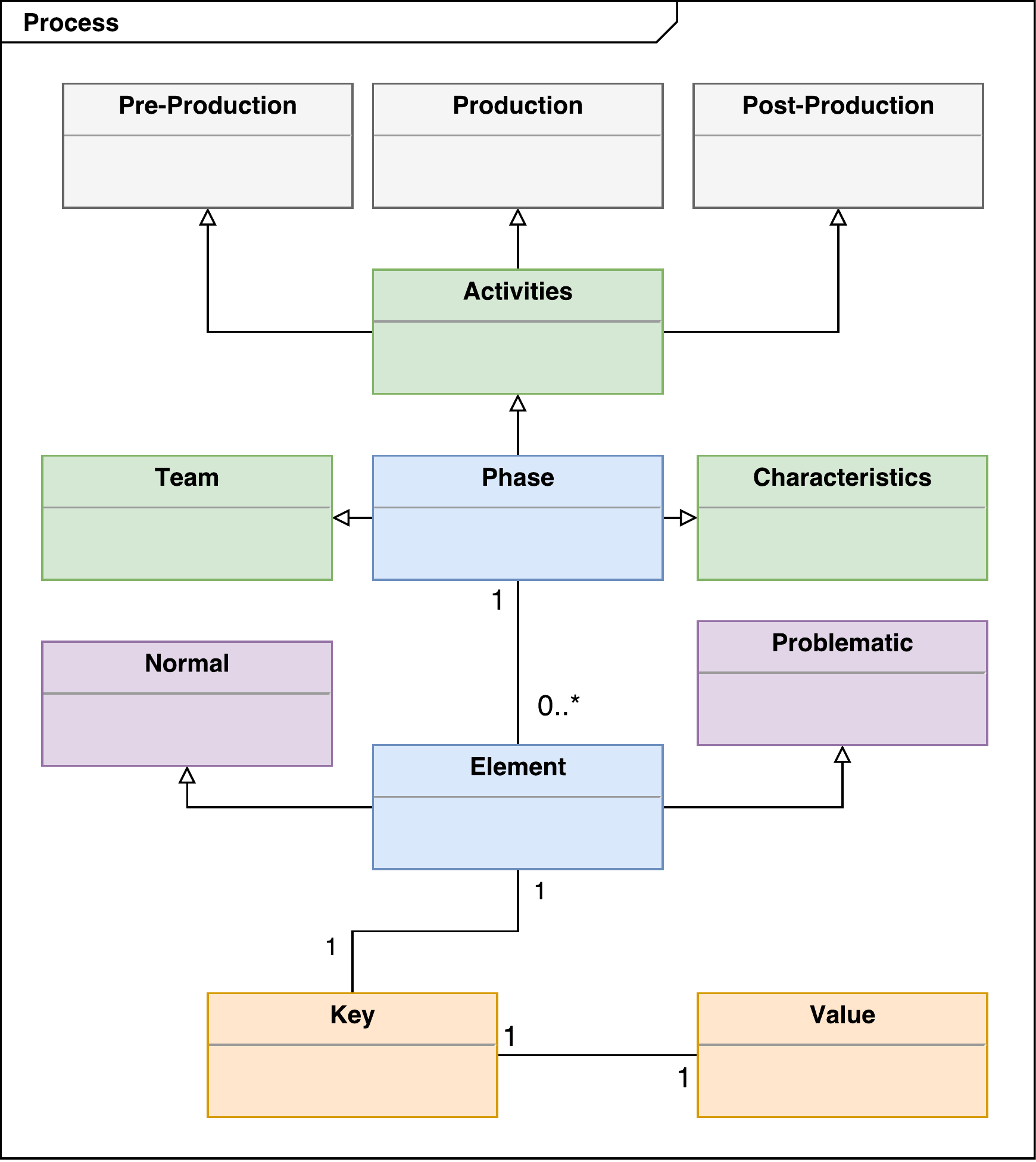}
    \caption{Architecture used to store process' elements.}
    \label{fig:class_diag_elements}
\end{figure}

We analyzed the data gathered from \textit{Gamasutra} by reading every postmortem, collecting information about the development processes, activities, and team characteristics.
We ignored details about game design or other information that does not pertain to the development processes.
We organized data---process' elements---in a database following the model described in Figure \ref{fig:class_diag_elements}.

We defined a development process as a set of elements separated in different phases and sub-phases. The \textbf{Phase} categorizes the elements of the process, which can be:

\begin{itemize}
\item \textbf{Activities}, which represent tasks or steps describing work units that result in artifacts. For example, the activity of \textit{prototyping}.

\item \textbf{Team}, which pertain to the characteristics of teams. For example, the team's characteristic of \textit{outsourcing}.

\item \textbf{Characteristics}, which describe meta-data about the development process. For example, the characteristic of \textit{project duration}.
\end{itemize}

Based on previous work \cite{bates2004game,moore2010game}, we define three main process sub-phases in which \textbf{Elements} in Activities can be contained:

\begin{itemize}
\item \textbf{Preproduction}, containing activities like \textit{brainstorming}, \textit{prototyping}, \textit{validation} and tasks to find the ``fun factor'';
\item \textbf{Production}, where technical tasks occur;
\item \textbf{Post-production}, containing any activities executed after the game has been launched.
\end{itemize}

Each Element can have one of two possible status: \textbf{Normal}, when the postmortem author does not report any problem regarding the activity execution, and \textbf{Problematic}, when developers encountered some problems related to the activity.

Each element has an index name, which we call \textbf{Key}, as well as a description, called \textbf{Value}. We put the postmortem extracted data in a map data-structure in which the \textit{key} is an element name and the \textit{value} is a quotation gathered from postmortems. Listing \ref{list:md} shows an example of data extracted from postmortems and stored in the map data-structure.

\begin{lstlisting}[caption=Example of a map data-structure used to store data from postmortems analysis., label=list:md, style=md]
# Slow Down, Bull
## team
* team
	* Besides me, the team for Slow Down, Bull was composed entirely of contractors (though some still had some (...)
## activities
* **initial prototyping**
	* Because the whole initial process was a bit of an experiment, we spent a long time with just me working on (...)
## characteristics
* expertise source
	* In a way, the act of consulting an expert became a form of delegation, and pure brain-expertise became a (...)
## feedback
* I would switch the position of initial prototyping and early play testing. The prototype had to exist before (...)
\end{lstlisting}

\begin{figure*}
	\centering
	\includegraphics[width=.8\linewidth]{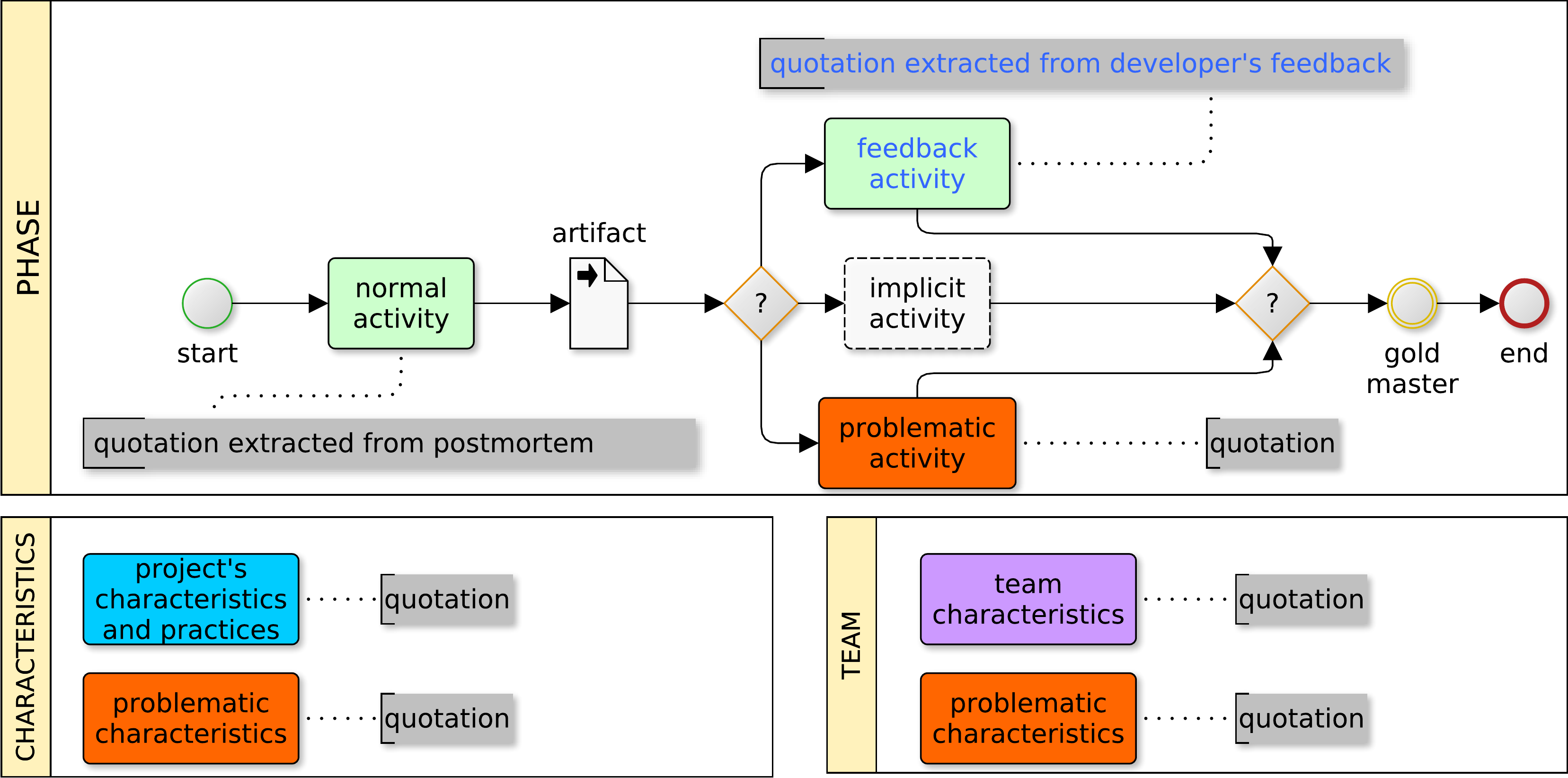}
    \caption{We use a visualization template (BPMN) to display uniformly processes graphically.}
	\label{fig:process-template}
\end{figure*}

For ease of conveying information from the processes to developers, we defined a process structure template file with the elements of process described earlier to generate a graphical representation of the processes. Figure~\ref{fig:process-template} shows the elements.

Finally, to improve the reliability of the \textbf{extracted processes}, we showed these processes to the postmortem authors and game developers, asking for feedback about these processes. We showed the extracted process and ask the following questions:

\begin{enumerate}
\item ``Is this process similar of what you used when developing your game?''
\item ``What are the elements that does not make sense?''
\item ``Would you add something important?''
\end{enumerate}

With their answers, we modified the extracted process according to their feedbacks. This step helps us to ensure that the extracted processes are as close as possible to the development processes that they used during the video game development.

\subsection{Capturing Context}
\label{sec:m2}

We encoded the data gathered from postmortems in JSON format and put it in a database to perform queries. The JSON structure consists of five \textit{key:value} for each process element. A process is thus a collection of \textit{N} elements. Listing~\ref{list:jsonstructure} shows an example of the JSON format.

\begin{lstlisting}[caption=JSON structure used to storage process' elements., label=list:jsonstructure, style=code]
{
	"game" : "Slow Down, Bull",
	"phase" : "activities",
	"element" : "exploration phase",
	"desc" : "We were able to iterate through a ton of different experiments, many of which were discarded failures, but which paved the path for the strongest mechanics in the game",
	"prob" : false
}
\end{lstlisting}

Although the data is structured, it is not normalized. Some elements have the same meaning and, therefore, must be merged to a common description. We create an \textbf{abstraction dictionary} that contains all the elements and their respective indexes. For example, the elements \textit{``local play testing''} and \textit{``beta testing''}, refer to the activity \textit{``testing''}.

Lastly, we must define the variables that will be used by the algorithm as project context. A project context defines its characteristics, which are the variables used by the algorithm. There exist some attempts to contextualize non-video game software projects \cite{kruchten2013contextualizing,Boehm:2003}. However, the sets of variables are larger in video game projects than in other projects because of their multidisciplinarity.

We divide the \textbf{video game context} into six categories, totalizing 61 variables: \textit{activities}, \textit{team}, \textit{management}, \textit{technical}, \textit{platform}, and \textit{design}. Table \ref{tab:context} shows all the variables of the video game context.

\begin{table}[!ht]
\footnotesize
\centering
\caption{Video Game context variables.}
\label{tab:context}
\begin{tabularx}{\linewidth}{@{}lll@{}}
\toprule
\#  & Group              & Description                              \\ \midrule
    v01 & Activities         & Agile                                    \\
    v02 &                    & Prototyping                              \\
    v03 &                    & Performance Optimization                 \\
    v04 &                    & Tools Development                        \\
    v05 &                    & Outsourcing: Assets                      \\
    v06 &                    & Outsourcing: Work                        \\
    v07 &                    & Pre-Production: Short                    \\
    v08 &                    & Pre-Production: Long                     \\
    v09 &                    & Post-Production: Normal                  \\
    v10 &                    & Post-Production: Heavy                   \\
    v11 &                    & Reuse: Code                              \\
    v12 &                    & Reuse: Assets                            \\
    v13 &                    & Testing: In-Ouse Qa Team                 \\
    v14 &                    & Testing: Closed Beta                     \\
    v15 &                    & Testing: Open Beta                       \\
    v16 &                    & Testing: Early Access                    \\
    v17 &                    & Marketing/Pr: Self                       \\
    v18 &                    & Marketing/Pr: Outsourced                 \\
    v19 & Team               & Size: \textless=5                        \\
    v20 &                    & Size: 5 – 25                             \\
    v21 &                    & Size: \textgreater25                     \\
    v22 &                    & Type: Single                             \\
    v23 &                    & Type: Collaborative                      \\
    v24 &                    & Distribuited                             \\
    v25 & Management         & Developer Type: First-Party              \\
    v26 &                    & Developer Type: Second-Party             \\
    v27 &                    & Developer Type: Third-Party              \\
    v28 &                    & Indie                                    \\
    v29 &                    & Funding: External                        \\
    v30 &                    & Funding: Self                            \\
    v31 &                    & Funding: Crowdfunding                    \\
    v32 &                    & Publisher: External                      \\
    v33 &                    & Publisher: Self (same Developer)         \\
    v34 & Technical          & Intelectual Property: Port               \\
    v35 &                    & Intelectual Property: Remaster / Reboot  \\
    v36 &                    & Intelectual Property: Franchise Sequence \\
    v37 &                    & Intelectual Property: Expansion          \\
    v38 &                    & Intelectual Property: Mod                \\
    v39 &                    & Intelectual Property: New Ip             \\
    v40 &                    & Engine: In-House (new)                   \\
    v41 &                    & Engine: In-House Ready                   \\
    v42 &                    & Engine: Off-The-Shelf                    \\
    v43 & Platform           & Console: Microsoft                       \\
    v44 &                    & Console: Sony                            \\
    v45 &                    & Console: Nintendo                        \\
    v46 &                    & Pc: Windows                              \\
    v47 &                    & Pc: Mac                                  \\
    v48 &                    & Pc: Linux                             \\
    v49 &                    & Mobile: Android                          \\
    v50 &                    & Mobile: Ios                              \\
    v51 & Design             & Genre: Action                            \\
    v52 &                    & Genre: Action-Adventure                  \\
    v53 &                    & Genre: Adventure                         \\
    v54 &                    & Genre: Role-Playing                      \\
    v55 &                    & Genre: Simulation                        \\
    v56 &                    & Genre: Strategy                          \\
    v57 &                    & Genre: Puzzle                            \\
    v58 &                    & Genre: Sports                            \\
    v59 &                    & Mode: Single-Player                      \\
    v60 &                    & Mode: Multi-Player (offline)             \\
    v61 &                    & Mode: Multi-Player Online                \\ \bottomrule
\end{tabularx}
\end{table}


\subsection{Producing Recommendations}
\label{sec:m3}

The project contexts obtained from different postmortems are input for \textit{Principal Component Analysis} (PCA), discussed in Section \ref{sec:b-pca}, to identify and quantify the relations in large sets of variables. Principal Component Analysis (PCA) was used in this research because it was well suited to the problem and its simplicity, enabling a quick prototyping. Others techniques were used, like \textit{Decision Trees} and \textit{K-Means} algorithms. However, given the \textit{labels} restrictions, we changed the approach.

The result of applying PCA on the project context is an array of similar projects. With this list, we query the database for all elements related to the projects and perform a merge, respecting the elements' phase.
Because we stored the process' elements without its relationship, that is, flow between the activities, after put all similar elements together, we had to allocate the elements manually.

\subsection{Presenting the Recommendations}
\label{sec:m4}

The last step is to present the recommended activities: the development process. We choose to display recommended processes graphically, using the same graphical representation used for the extracted processes. Consequently, we use the same visualization template (see Figure \ref{fig:process-template}).


\section{Recommendation System Development}
\label{sec:dev}

We now present the recommendation system and how we built it. We followed the same four steps as in Section~\ref{sec:rmethod}.

\subsection{Data Preprocessing}
\label{sec:d1}

We used a Web scrapping tool to gather postmortems from \textit{Gamasutra} automatically. \textit{Gamasutra} was our source of information of choice because of the size of its postmortems database and its reliability.
We check and filtered data, extracting the details of the video game projects, like game title, author(s), release date, url, as well as data from \textit{Wikipedia}, like game mode, genre, developers, engine, etc. The total amount of game projects was 234 from 1997 (which is the oldest postmortem we found) to 2016.

\begin{figure}[ht]
	\centering
	\includegraphics[width=1\linewidth]{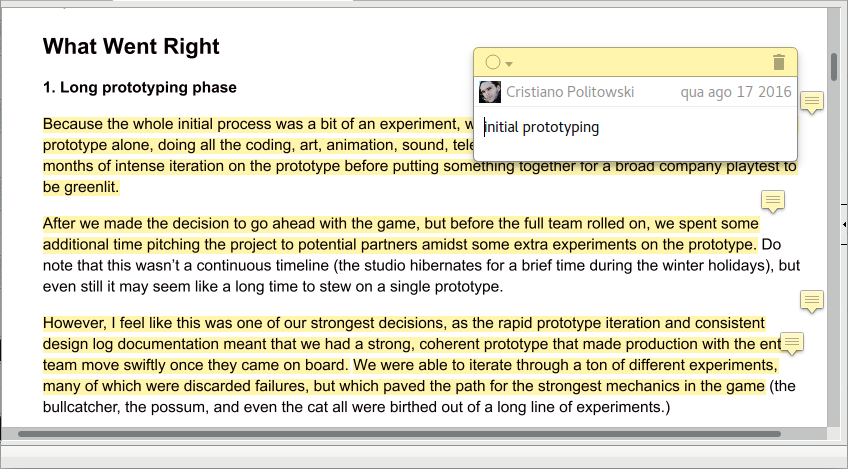}
	\caption{Example of the postmortem analysis made to extract and index the relevant data about the video games.}
	\label{fig:mendeley-example}
\end{figure}

Starting from the most recent postmortem, we analyzed and extracted the data to populate our JSON database.
To analyze the postmortems we made use of \textit{Mendeley} and its feature to create notes. Every information regarding the development process was highlighted and tagged with an index. For example, Figure~\ref{fig:mendeley-example} shows a piece of a postmortem in which we highlighted some excerpts and associated a index with each one, like \textit{``initial prototyping''}.

Some postmortems do not have useful data and, thus, among the 100 postmortems that we analyzed we kept only 55 that had sufficient information to create a development process. The other postmortems mainly focused on game designs or gave too little details about team and the project itself.

We use \textit{GraphViz} to display processes extracted from the notes graphically. \textit{GraphViz} uses a special text format to generate graphical files, which can be saved as PDF and PNG files. We created a script that query the database, which contains the process elements, and generate a new file, using the process template (see Figure \ref{fig:process-template}), in DOT syntax, which can be converted into an image.

We requested feedback from the developers, authors of the postmortems, about the extracted processes. We sent a on-line form asking about the completeness and viability of the processes. A total of 20 forms were sent with 7 answers (33\% answer rate). We refactored few wrong or misplaced elements in some of the processes. All of the answers agreed with the extracted processes, for example in the citation of the postmortem author, developer of the game ``Prune'':

\begin{quotation}
\noindent
\textit{``Yeah, that's about right. I would maybe call the first 6 months of work on the game ``preproduction'', which included things like prototyping, task prioritization, and play testing.''}
\end{quotation}

\begin{sidewaysfigure*}
	\centering
	\includegraphics[width=1\linewidth]{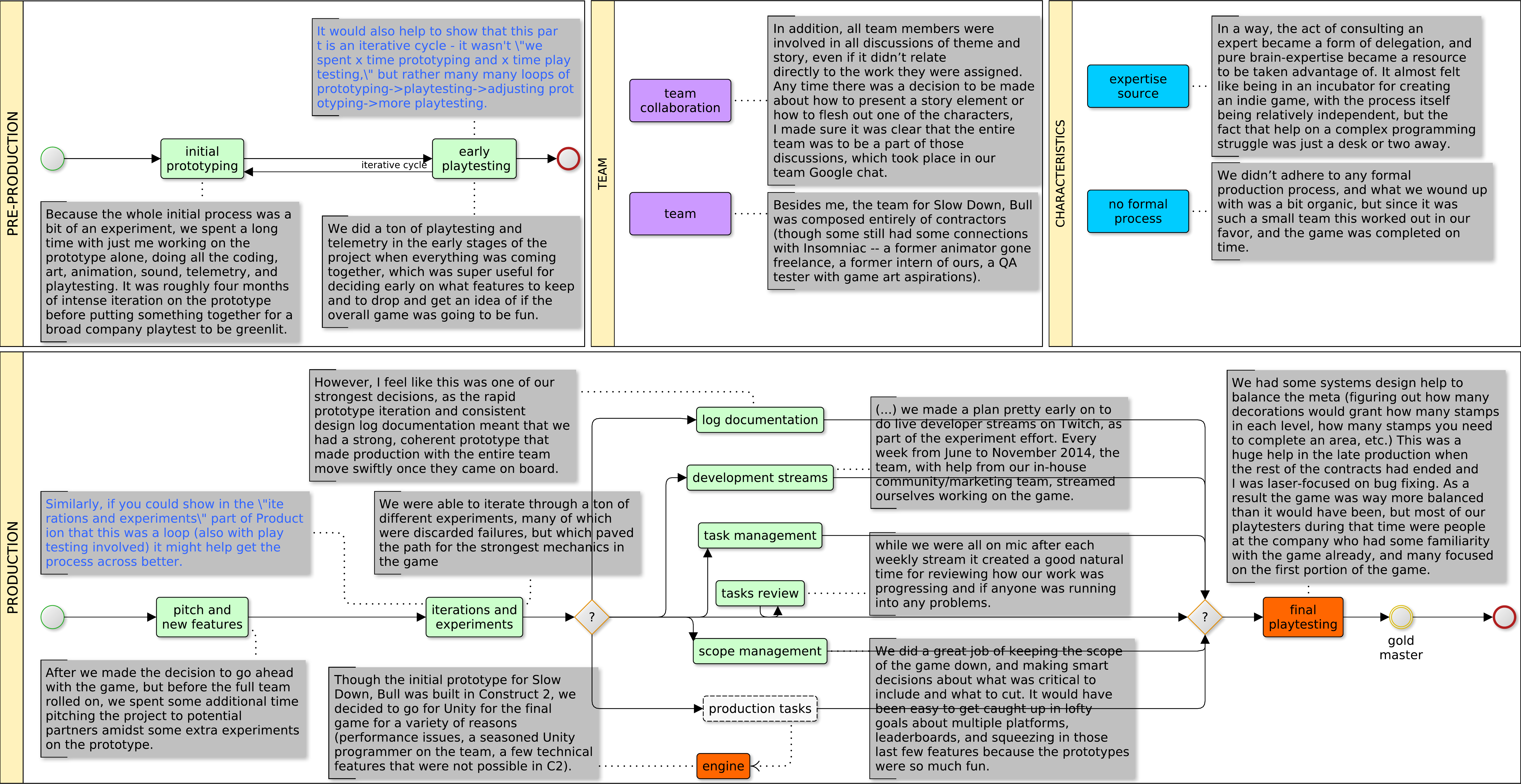}
    \caption{Extracted development process from \textit{Slow down bull} game.}
	\label{fig:extracted-process}
\end{sidewaysfigure*}

Figure \ref{fig:extracted-process} shows an example of an extracted process, validated by the authors of the origin postmortem, of the game project \textit{Slow down, Bull}\footnote{To save space, we manually redesigned the process, respecting the template rules.}. This process shows two phases: preproduction and production, without post-production. Each phase has a start and an end date, demarcated by the dark discs. In both phases, the elements in green are activities that occurred naturally during the development while the red ones encountered problems.

Considering that we extracted process elements and, normally, they do not had chronological information about its execution, we cannot, precisely describe the process flow. In these cases, we marked such elements with two question marks (``?''). In production, the element described as ``gold'' indicates the final version of the game, the complete game. Each activity, in the gray boxes, has a direct related to the postmortem. There are two frames containing the characteristics and general practices together with the ones related to the development team.

\subsection{Capturing Context}
\label{sec:d2}

We normalized all the elements keys using the \textbf{abstraction dictionary} and inserted the elements in the database. We chose \textit{MongoDB}\footnote{https://www.mongodb.com/} as storage engine because it was easy to import JSON files. We stored a total of \textbf{913 different elements} in the database from \textbf{55 projects}.

With the data ready, we now needed a \textit{project context} (see Table~\ref{tab:context}). We analyzed the project data and, according to their characteristics, manually build contexts. The context values (see Table~\ref{tab:context_values}) are composed of 61 variables ([v01..vN]) from the 55 extracted process (game projects, [g1..gN]). Every time a new project is added to the database, it helps in future recommendations.

\begin{table}[!ht]
\footnotesize
\centering
\caption{Context table filled (partialy).}
\label{tab:context_values}
\begin{tabularx}{\linewidth}{@{}Xrrrrrrrrrr@{}}
\toprule
    & g1 & g2 & g3 & g4 & g5 & g6 & g7 & g8 & g9 & gN \\ \midrule
v01 & 1  & 1  & 1  & 1  & 1  & 1  & 1  & 1  & 1  & …  \\
v02 & 0  & 1  & 1  & 1  & 1  & 1  & 0  & 0  & 1  & …  \\
v03 & 0  & 0  & 1  & 0  & 0  & 0  & 0  & 0  & 0  & …  \\
v04 & 0  & 0  & 0  & 0  & 0  & 0  & 0  & 0  & 0  & …  \\
v05 & 0  & 0  & 0  & 0  & 0  & 0  & 1  & 0  & 0  & …  \\
v06 & 1  & 0  & 0  & 0  & 0  & 1  & 0  & 0  & 1  & …  \\
v07 & 0  & 1  & 1  & 1  & 1  & 1  & 0  & 1  & 1  & …  \\
v08 & 1  & 0  & 0  & 0  & 0  & 0  & 1  & 0  & 0  & …  \\
v09 & 1  & 1  & 0  & 0  & 1  & 1  & 0  & 1  & 1  & …  \\
vN  & …  & …  & …  & …  & …  & …  & …  & …  & …  & …  \\ \bottomrule
\end{tabularx}
\end{table}

\subsection{Producing Recommendations}
\label{sec:d3}

\begin{figure}[ht]
	\centering
	\includegraphics[width=1\linewidth]{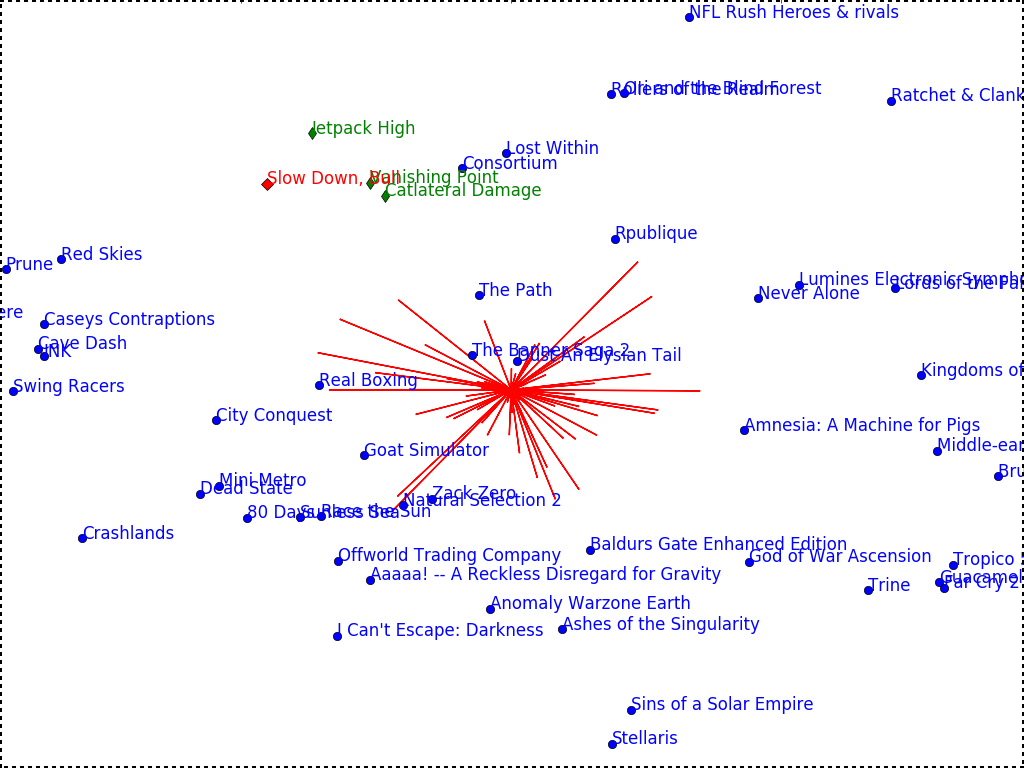}
	\caption{PCA  biplot using the game \textit{Slow down, Bull} as input project context.}
	\label{fig:pcaslow}
\end{figure}

With the projects contexts, we ran a script that uses \textit{PCA biplot} to show the samples' variability through a graph, that is, a graph showing the similarity of the 55 projects analyzed. Figure \ref{fig:pcaslow} shows the \textit{biplot PCA}, which is a representation of the correlation between the samples. The closest the dots, the more similar the projects. The input project context is from the game \textit{Slow down, Bull}, presented in red; the green dots show projects that are more similar, such as \textit{Jetpack High}, \textit{Vanishing Point}, and \textit{Catlateral Damage}.

With a list of similar projects, we queried the database for their elements. We then joined the elements resulting in a new process. We assembled the elements in a process manually, because we did not stored the processes flows.



\subsection{Presenting the Recommendations}
\label{sec:d4}

We used again \textit{Graphiz} with the visualization template to display the new recommended process (see Figure \ref{fig:process-template}). We converted the new recommended process in a textual DOT file, which can be compiled into an image file.

\begin{figure*}
	\centering
	\includegraphics[width=.8\linewidth]{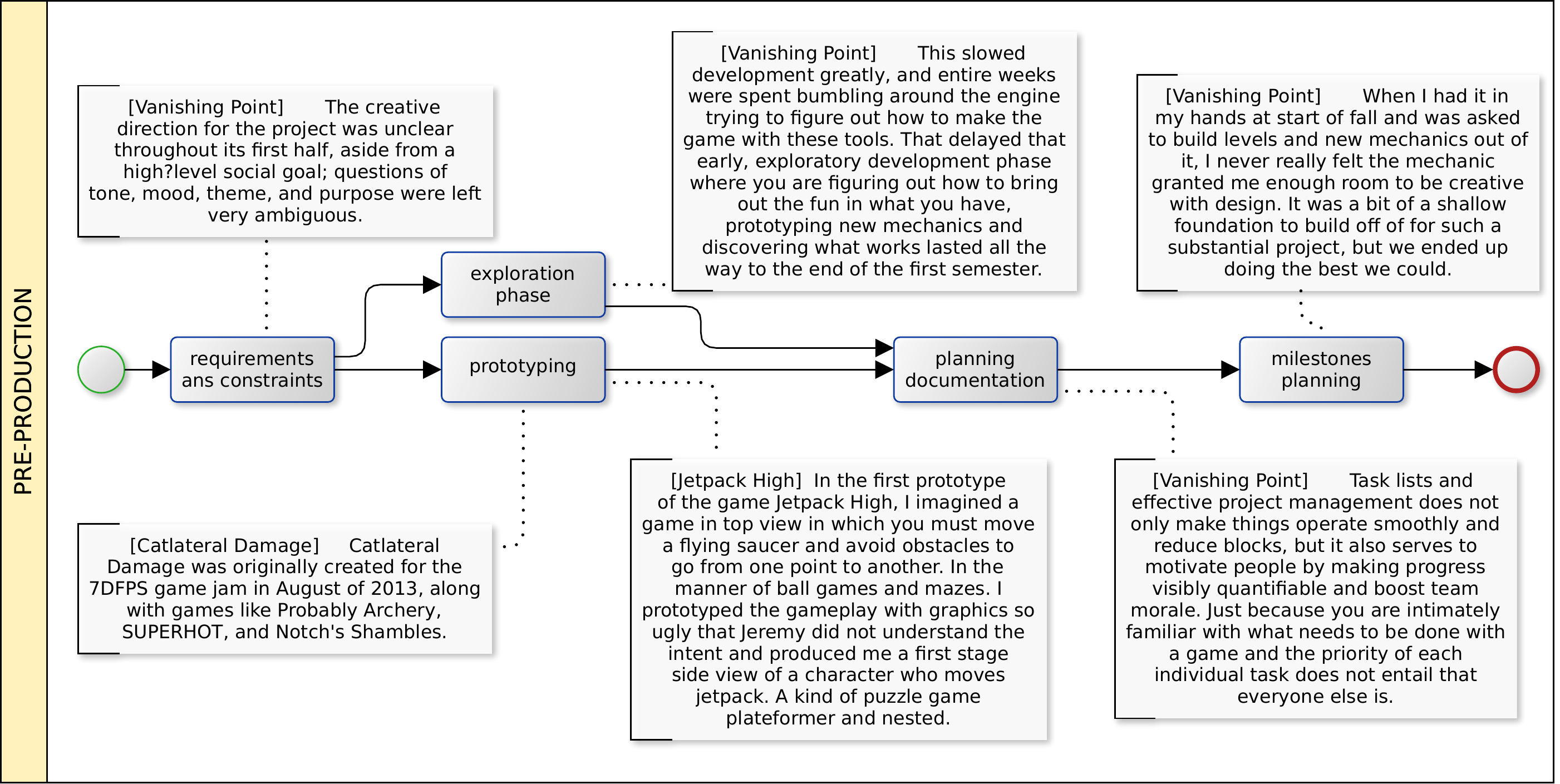}
	\caption{Preproduction phase of the recommended process for the game \textit{Slow down, Bull}.}
	\label{fig:slownewiterativebpmnpre}
\end{figure*}

\begin{sidewaysfigure*}
	\centering
	\includegraphics[width=1\textwidth]{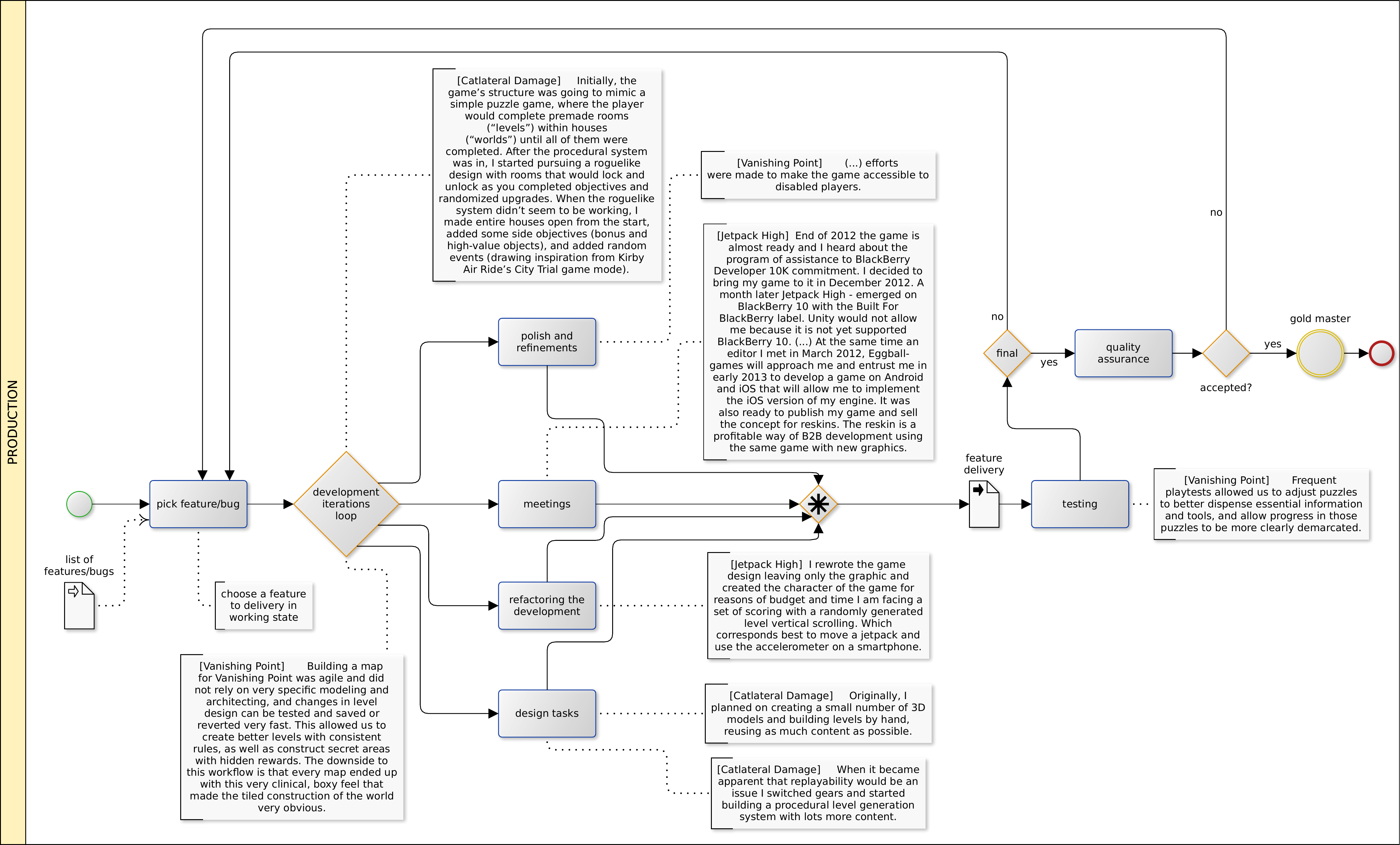}
    \caption{Recommended process production phase for the game \textit{Slow down, Bull}.}
	\label{fig:slownewiterativebpmnpro}
\end{sidewaysfigure*}

Figure \ref{fig:slownewiterativebpmnpre} shows an example of a recommended process and its preproduction phase while Figure \ref{fig:slownewiterativebpmnpro} shows the production phase using the game context of the game \textit{Slow down, Bull} shown in Table \ref{tab:contextslow}.

\begin{table}[!ht]
\footnotesize
\centering
\caption{Project context of the game \textit{Slow down, Bull}. The \textit{false} values were omitted.}
\label{tab:contextslow}
\begin{tabularx}{\linewidth}{@{}lXr@{}}
\toprule
\#  & Context                          & g10 values \\ \midrule
p01 & Agile                            & 1   \\
p02 & Prototyping                      & 1   \\
p06 & Outsourcing: Work                & 1   \\
p08 & Pre-Production: Long             & 1   \\
p09 & Post-Production: Normal          & 1   \\
p13 & Testing: In-Ouse Qa Team         & 1   \\
p19 & Size: \textless=5                & 1   \\
p22 & Type: Single                     & 1   \\
p24 & Distribuited                     & 1   \\
p27 & Developer Type: Third-Party      & 1   \\
p28 & Indie                            & 1   \\
p29 & Funding: External                & 1   \\
p33 & Publisher: Self (same Developer) & 1   \\
p39 & Intelectual Property: New Ip     & 1   \\
p42 & Engine: Off-The-Shelf            & 1   \\
p46 & Pc: Windows                      & 1   \\
p57 & Genre: Puzzle                    & 1   \\
p59 & Mode: Single-Player              & 1   \\ \bottomrule
\end{tabularx}
\end{table}

Figure \ref{fig:slownewiterativebpmnpre} shows the preproductiona and has the activities: \textit{requirements and constraints}, \textit{exploration phase}, \textit{planning documentation}, and \textit{milestones planning} linked with quotations from the game \textit{Vanishing Point}; \textit{prototyping} is linked to the games \textit{Catlateral Damage} and \textit{Jetpack High}.

Figure \ref{fig:slownewiterativebpmnpro} shows the production phase. The process flow starts with an activity to choose a feature or a bug, depending on the iteration type (component delivery or fixing a problem). As input, this activity receives a list of game properties from the preproduction phase. This list is similar to \textit{sprint backlog} used in agile methods like \textit{Scrum} \cite{schwaber2002agile}.

The process flow continues with the \textit{development iterations loop}. This element \textit{development iterations loop} marks the beginning of the productions activities. Two quotes, gathered from the postmortems of games \textit{Vanishing Point} and \textit{Catlateral Damage} are highlighted. The premise is similar to \textit{sprints} in agile methods in which the process occurs in short cycles with continuous deliveries of complete features. At the end of each iteration, a new, complete component must be aggregated to the existent game. The artifact \textit{feature delivery} represents this component.

The technical activities happen inside each iteration. There are four described activities, all of them connected with at least one citation from a game: \textit{polish and refinements}, \textit{meetings}, \textit{refactoring the development}, and \textit{design tasks}.

According to the development flow of the generated process, after the delivery loop  of components, the \textit{testing} activity will start, which is crucial to the success or failure of a game. This activity may happen in many ways, for example, the team can have a play testing laboratory with users, a specialized testing team, automated tests, users testing, beta testing, among others. Oonly one project, \textit{Vanishing Point} mentions explicitly testing.

After the end of the iterations, with the game tested, the flow arrives on a condition to test whether all the iterations are completed. If positive, the next activity is \textit{quality assurance} that suggests to have the game reviewed by an external entity, either a publisher or a platform (\textit{Sony}, \textit{Steam}, \textit{Microsoft}) to identify possible nonconformities with standards required by such publisher or platform. Then, the final \textit{build} is realized, fulfilling the last requirements, to obtain and publish a \textit{gold master}.

\section{Evaluation}
\label{sec:validation}

To assess our recommendation system, we performed three types of evaluations:
\begin{enumerate}[label=\Alph*.]
\item \textbf{Quantitative}, by measuring Correctness and Coverage of the recommended elements in Sections~\ref{sec:metval1};
\item \textbf{Qualitative}, by asking feedback from the game developers on the recommended process in Sections~\ref{sec:metval2};
\item \textbf{Case Study}, by applying a recommended process in a real world project in Section~\ref{sec:mcs}.
\end{enumerate}

\subsection{Quantitative Evaluation}
\label{sec:metval1}

To quantitatively assess the recommendation system, we chose four projects which we extracted from postmortems and validated it with their respective developers. We extracted their contexts and generated processes using our recommendation system. We compared if each recommended element is present in the extracted (validated) process. We used four measures to asses our recommendation system quantitatively \cite{avazpour2014dimensions}, shown in Table~\ref{tab:dimensions_quanti}.

\begin{table}[!ht]
	\footnotesize
	\centering
	\caption{Recommendation-centric dimensions.}
	\label{tab:dimensions_quanti}
	\begin{tabularx}{\linewidth}{@{}lX@{}}
		\toprule
		Dimensions  & Description \\
        \midrule
		Correctness & How close are the recommendations to a set of recommendations that are assumed to be correct? \\
		Coverage    & To what extent does the recommendation system cover a set of items or user space? \\
		Diversity   & How diverse (dissimilar) are the recommended items in a list? \\
		Confidence  & How confident is the recommendation system in its recommendations? \\
        \bottomrule
	\end{tabularx}
\end{table}

Recommendation-centric dimensions primarily assess the recommendations generated by the recommendation system itself: their coverage, correctness, diversity, and level of confidence \cite{avazpour2014dimensions}. We chose two metrics to evaluate our recommendation system: \textbf{Correctness} and \textbf{Coverage}. \textit{Diversity} does not apply to our system because our dataset is composed by unique elements: each element stored in the database is unique. Also, \textit{confidence} requires observable variables with users and we received feedback from developers, as in Section~\ref{sec:metval2}.

\subsubsection{Correctness Metric}

The objective of the correctness metric is to assess how useful are the results without being overwhelming with unwanted items \cite{avazpour2014dimensions}. With this metric, we measure \textbf{how close the recommended process is compared to the expected, ideal process}, and already validated process.

There are different means of evaluating the recommended processes, like \textit{Predicting User Ratings} and \textit{Ranking Items}. However, we chose \textbf{Recommending Interesting Items}, which tests if the recommendation system provide process elements that users may like to use.

We fed the recommender system with data from a project context which project we already validated, a process extracted from a postmortem verified by its authors and--or developers. The confusion matrix shown in Table \ref{tab:confusion-matrix} compares each process element of the \textbf{recommended process} test set with the elements from the \textbf{extracted process} (training set).

\begin{table}[!ht]
	\footnotesize
	\centering
	\caption{Confusion matrix used to calculate correctness \cite{avazpour2014dimensions}.}
	\label{tab:confusion-matrix}
	\begin{tabularx}{\linewidth}{@{}Xll@{}}
		\toprule
		& Recommended          & Not Recommended       \\ \midrule
		Used           & True Positives (TP)  & False Negatives (FN)  \\
		Not Used & False Positives (FP) & True Negatives (TN)    \\
		 \bottomrule
	\end{tabularx}
\end{table}

For example, consider a recommended process with an element \textit{team} and an extracted process that contains this element, we have a true positive (TP) element. Otherwise, if the recommended element is not contained in the extracted process, then we have a false positive (FN) element. Then, we benchmarked our recommendation system by calculating some metrics\footnote{Except Specificity, which is a heuristic measure, all other values are in [0..1].}:

\begin{itemize}
\item \textbf{Precision}: the percentage of the elements predicted to be relevant that are actually relevant. \[ precision = \frac{TP}{TP + FP} \];

\item \textbf{Recall}: the percentage of the elements that are actually relevant that are predicted to be relevant. \[ recall = \frac{TP}{TP + FN} \]

\item \textbf{Accuracy}: the percentage of the elements available that are either correctly recommended or correctly not recommended. \[ accuracy = \frac{TP + TN}{TP + TN + FP + FN} \]

\item \textbf{False positive rate}: the percentage of the irrelevant elements that are predicted to be relevant. \[ falsePositiveRate = \frac{FP}{FP + TN} \]

\item \textbf{False negative rate}: the percentage of the elements predicted to be irrelevant that are actually relevant. \[ falseNegativeRate = \frac{FN}{FN + TN} \]

\item \textbf{Specificity}: a heuristic measure \cite{robillard2008topology} of the likelihood of the relevance of an element given its relation to other elements, some of which are known to be relevant. \[ specificity = \frac{TN}{FP + TN} \]

\item \textbf{F-measure}: a measure combining precision and recall. \[ F-measure = 2 x \frac{precision X recall}{precision + recall} \]
\end{itemize}

\subsubsection{Coverage Metric}

Coverage refers to the proportion of available information for which recommendations can be made. It can be calculated as catalog coverage (about elements) or prediction coverage (about developers) \cite{avazpour2014dimensions}. We use the former because we want to assess the recommended processes. This metric can be calculated as the proportion of elements (process elements) that we can be recommended:

\[ catalogCoverage =  \frac{\left |  Sr \right |}{\left |  Sa \right |} \]

The elements available for recommendation may not all be of interest to a developer so we used a previous approach \cite{ge2010beyond}, which uses weighted catalog coverage for balancing the decrease in coverage by usefulness for developers:

\[ weightedCatalogCoverage =  \frac{\left |  Sr \cap Ss \right |}{\left |  Ss \right |} \]

\noindent where:

\begin{itemize}
	\item \textit{Sr}: is set of process elements recommended;
	\item \textit{Sa}: is the set of all available elements;
	\item \textit{Ss}: is the set of elements useful to developers (validated elements).
\end{itemize}

\begin{figure}[ht]
	\includegraphics[width=\linewidth]{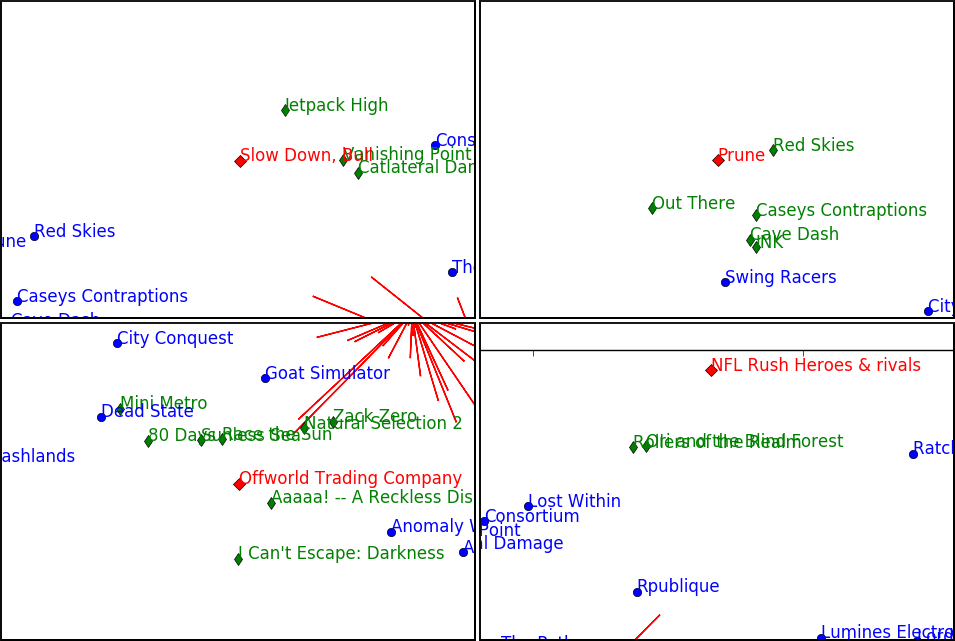}
	\caption{\textit{Biplots PCAs} from analyzed projects showing their similarity. From left to right: \#1, \#2, \#3, \#4.}
	\label{fig:b}
\end{figure}

We computed correctness and coverage of the recommendation system by comparing the similarity between the \textbf{extracted} processes, validated by their authors, and the \textbf{recommended} processes, generated by the recommendation system. We used four different project contexts, resulting in an array of similar projects. Figure \ref{fig:b} shows the resulting \textit{byplots}.

\begin{enumerate}
\item The game project \textbf{Prune} is similar to \textit{Red Skies, Out There, INK, Cave Dash, Caseys Contraptions};
\item The game project \textbf{Offworld Trading Company} is similar to \textit{Mini Metro, Sunless Sea, Race the Sun, I Cant Escape: Darkness, 80 Days, Natural Selection 2, Aaaaa! -- A Reckless Disregard for Gravity, Zack Zero};
\item The game project \textbf{NFL Rush Heroes \& rivals} is similar to \textit{Ori and the Blind Forest, Rollers of the Realm};
\item The game project \textbf{Slow Down, Bull} is similar to \textit{Catlateral Damage, Jetpack High, Vanishing Point}.
\end{enumerate}

We created a script to compare each element from each recommended process, simplified in the pseudo-code listed in Algorithm \ref{list:compare_elements}. We compared if the recommended elements were in the extracted processes, which we considered as correct, because they were reviewed by the postmortems authors.
If the elements recommended by the recommendation system were in the extracted processes, then they received a \textit{True Positive (TP)}.
Otherwise, if the recommended elements were not part of the extracted processes, then they received a \textit{False Positive (FP)}\footnote{Even if an element is not part of the extracted process, it still can make sense and be useful for the developers \cite{shani2011evaluating}.}.
If process elements were in the extracted processes but not recommended, they received a \textit{False Negative (FN)}.
Finally, if process elements were not part of the recommended process nor in the extracted process, then they received \textit{True Negative (TN)}.

\begin{algorithm}
\footnotesize
\caption{Pseudo-code to compare the process elements.}
\label{list:compare_elements}
\begin{algorithmic}[1]
	\For{$gameTested \in gamesTested[]$}
		\State $similarGames[] \gets getSimilarGamesPCA(gameTested)$
		\State $elemGameTested[] \gets getAllElements(gameTeste)$
		\State $elemRecommended[] \gets getAllElements(similarGames)$
		\State $elemNotRecommended[] \gets getAllElementsExcept(elemGameTested + elemRecommended)$

		\If{$elemRecommended[i] \in elemGameTested[]$}
			\State $truePositive[] \gets elemRecommended[i]$
		\Else
			\State $falsePositive[] \gets elemRecommended[i]$
		\EndIf

		\If{$elemNotRecommended[i] \in elemGameTested[]$}
			\State $falseNegative[] \gets elemNotRecommended[i]$
		\Else
			\State $trueNegative[] \gets elemNotRecommended[i]$
		\EndIf

	\EndFor
\end{algorithmic}
\end{algorithm}

Table~\ref{tab:confmatrix_values} shows the confusion matrix with aggregated data and Table~\ref{tab:validation_values} shows the metrics results.
The high number of \textit{false negatives} and low number of \textit{true positives} can be improved by increasing the range of similarity: by adding more similar projects. Adding more projects would also increase \textit{recall}, given that recall was low. However, allowing for a longer list of recommendations improves \textit{recall} but is likely to reduce \textit{precision} and improving \textit{precision} may decrease recall \cite{salfner2010survey}.

\begin{table}[!ht]
	\footnotesize
	\centering
	\caption{Confusion matrix with aggregated data.}
	\label{tab:confmatrix_values}
	\begin{tabularx}{\linewidth}{@{}Xrrrr@{}}
      \toprule
          & T. Positive & F. Positive & F. Negative & T. Negative \\ \midrule
      \#1 & 32 & 44 & 285 & 540 \\
      \#2 & 23 & 105 & 153 & 620 \\
      \#3 & 10 & 32 & 171 & 692 \\
      \#4 & 14 & 19 & 358 & 505 \\ \bottomrule
	\end{tabularx}
\end{table}

Although \textit{accuracy} and \textit{specificity} rates were good, all others values were low. The recommendation system reached a precision above 40\% in two projects. We used \textit{PCA} to find similarity and will consider other algorithms in future work. We can also improve our recommendation system by improving the context variables and samples number.

Table \ref{tab:validation_values} shows the \textit{coverage} metric results with the low values in \textit{Catalog} and \textit{Weighted Catalog}, which could be improved if more elements where recommended. Increasing \textit{coverage} may increase \textit{correctness} as well.

\begin{table*}
	\footnotesize
    \centering
    \caption{Correctness and coverage results.}
    \label{tab:validation_values}
    \begin{tabularx}{\linewidth}{@{}Xrrrrrrrrrrrr@{}}
    \toprule
      & \multicolumn{7}{l}{\textbf{Correctness}} & \multicolumn{5}{l}{\textbf{Coverage}} \\ \midrule
      & Precision & Recall & Accuracy & FP Rate & FN Rate & Specificity & F-Measure & Sr & Sa & Ss & Catalog & W. Catalog \\ \midrule
    \#1 & 42,11\% & 10,09\% & 63,49\% & 7,53\% & 34,55\% & 92,47\% & 16,28\% & 76 & 913 & 317 & 8,32\% & 10,09\% \\
    \#2 & 17,97\% & 13,07\% & 71,37\% & 14,48\% & 19,79\% & 85,52\% & 15,13\% & 128 & 913 & 176 & 14,02\% & 13,07\% \\
    \#3 & 23,81\% & 5,52\% & 77,57\% & 4,42\% & 19,81\% & 95,58\% & 8,97\% & 42 & 913 & 181 & 4,60\% & 5,52\% \\
    \#4 & 42,42\% & 3,76\% & 57,92\% & 3,63\% & 41,48\% & 96,37\% & 6,91\% & 33 & 913 & 372 & 3,61\% & 3,76\% \\ \bottomrule
    \end{tabularx}
\end{table*}

\subsection{Qualitative Evaluation}
\label{sec:metval2}

We performed the qualitative validation using a set of questions in a on-line form that we sent to project authors and--or developers. For each one of the chosen project, that is, project \#1 to \#4 already used in the quantitative validation, we asked recipients to analyze both the \textbf{extracted} and the \textbf{recommended} processes. We also asked them to highlight the elements that were wrong or misplaced. Then, we asked then about the viability of the recommended process:

\begin{enumerate}
	\item ``In general, is this new workflow similar to what you used developing the game?''
	\item ``Which process elements does not make sense?''
	\item ``Do you add something important to this workflow?''
	\item ``If you began a new video game project similar to [GAME], what would be the feasibility of using this new process? (answer 5 to ``more feasible'' and 1 to ``unfeasible'')''
\end{enumerate}

In general, \textbf{all respondents agreed with the recommended processes}. They found that the processes were similar to the processes that they used during their game development. One developer stated that the activities flow is similar to what was actually done:

\begin{quotation}
\noindent
``Yes, this workflow looks similar to what we did with \textit{[game title]}, in that we did a lot of iteration with our feature development.''
\end{quotation}

In another example, the postmortem author and game developer answered about the generated process viability, stating that the most useful contribution may be during the project planning, but not for creation of any production pipeline.

\begin{quotation}
\noindent
\textit{``(...) this might be a useful thing to look at in the beginning, but I would not use it to create a production pipeline because the circumstances have almost assuredly changed since.''}
\end{quotation}

Still, the postmortem' author pointed out that the recommended process is interesting for analysis, but not something to be used practically to build a set of production activities. The developers claimed that each project demands a different process based on the game design choice.

\begin{quotation}
\noindent
\textit{``I think the tool that you describe has some really interesting implications from an analytical standpoint, but it is not something I would use in a practical sense, like I wouldn't use it to develop processes or pipelines from. In practice, every game requires different processes to make, and in fact the process used to make a game is as much a part of its design as the actual game. (...) The reasons for variation in process are vast and constantly changing. Even in big studios you tend to end up with processes that are perfect for making the game you were working on the year before (...) However, I still think it is valid work from an analytical standpoint, like as something to consider when having the conversation about dev process for the next game.''}
\end{quotation}

\subsection{Case Study}
\label{sec:mcs}

We performed a case study using qualitative metrics from user-centric dimensions to assess if the recommendation system fulfills the needs of developers \cite{avazpour2014dimensions}. We use the metrics in Table \ref{tab:dimensions_user}.

\begin{table}[!ht]
\footnotesize
\centering
\caption{Dimensions: User-centric.}
\label{tab:dimensions_user}
\begin{tabularx}{\linewidth}{@{}lX@{}}
\toprule
Dimensions      & Description \\
\midrule
Trustworthiness & How trustworthy are the recommendations? \\
Novelty         & How successful is the recommendation system in recommending items that are new or unknown to users? \\
Serendipity     & To what extent has the system succeeded in providing surprising yet beneficial recommendations? \\
Utility         & What is the value gained from this recommendation for users? \\
Risk            & How much user risk is associated in accepting each recommendation? \\
\midrule
\end{tabularx}
\end{table}

We interviewed with an team that is currently developing a game. We asked questions to measure these user-centric dimensions. Firstly, we gathered the project context and generated a new process using the recommendations system. Then, we presented the recommended processes to the team members and asked them to answer the following statements using a five-point Likert scale:

\begin{enumerate}
	\item ``(Trustworthiness) The recommendation is similar compared to my project.''
	\item ``(Novelty) The recommendation is new to me (regardless its usefulness).''
	\item ``(Serendipity) The recommendation is surprisingly good for my project.''
	\item ``(Utility) The recommendation is useful to my project.''
    \item ``(Risk) It would be risky to use the recommendation in my project (considering other practices already settled).''
\end{enumerate}

We chose a simulation-like video game project that uses \textit{Unity} engine and has been in development for more than three years. We asked the lead developers to fill the context form according to the project context, shown in Table~\ref{tab:contextcasestudy}.

\begin{table}[!ht]
\footnotesize
\centering
\caption{Context values from the project used in case study. The
\textit{false} values were omitted.}
\label{tab:contextcasestudy}
\begin{tabularx}{\linewidth}{@{}lXr@{}}
\toprule
\#  & Context                         & Values \\ \midrule
	v01 & Agile                           & 1      \\
	v02 & Prototyping                     & 1      \\
	v03 & Performance Optimization        & 1      \\
	v04 & Tools Development               & 1      \\
	v05 & Outsourcing Assets              & 1      \\
	v07 & Pre-Production Short            & 1      \\
	v11 & Reuse Code                      & 1      \\
	v12 & Reuse Assets                    & 1      \\
	v13 & Testing  In-Ouse Qa Team        & 1      \\
	v20 & Size 5 - 25                     & 1      \\
	v23 & Type Collaborative              & 1      \\
	v27 & Developer Type  Third-Party     & 1      \\
	v30 & Funding Self                    & 1      \\
	v33 & Publisher Self (same Developer) & 1      \\
	v39 & Intelectual Property New Ip     & 1      \\
	v42 & Engine Off-The-Shelf            & 1      \\
	v46 & Pc Windows                      & 1      \\
	v55 & Genre Simulation                & 1      \\
	v56 & Genre Strategy                  & 1      \\
	v59 & Mode Single-Player              & 1      \\ \bottomrule
\end{tabularx}
\end{table}

We appended the context values in the context table, executed the recommendation system, and obtained a list of similar projects: \textit{Ashes of the Singularity, Baldurs Gate Enhanced Edition, Natural Selection 2, Anomaly Warzone Earth and Aaaaa! -- A Reckless Disregard for Gravity.}

\begin{figure}[ht]
    \includegraphics[width=1\linewidth]{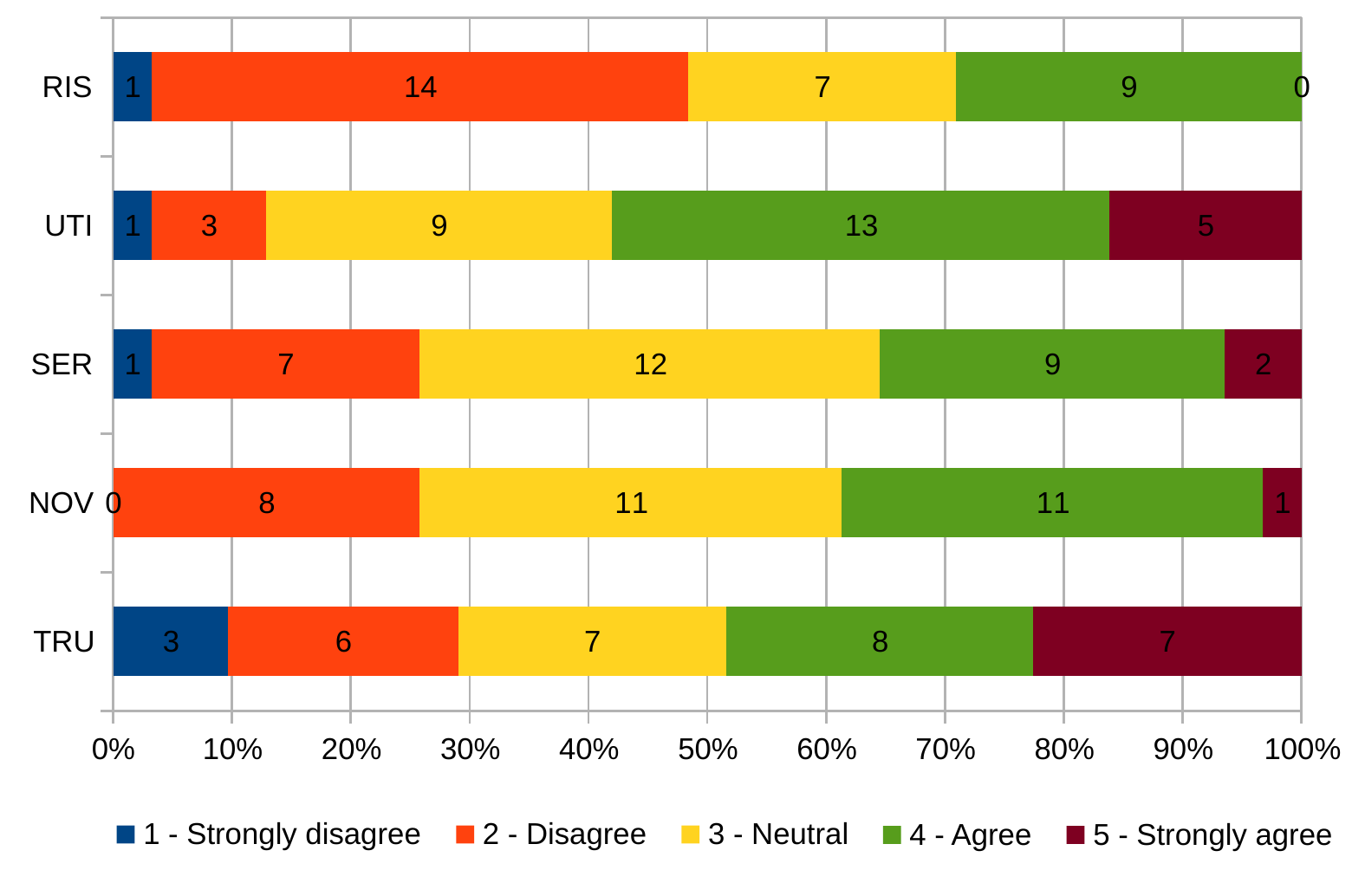}
    \caption{Case study results.}
    \label{fig:cs_results}
\end{figure}

A total of \textbf{31 elements} were recommended and so, for each one of them, we asked the development team to fill the statements, comparing the recommended elements with their actual activities. Table \ref{tab:cs_rec_elements} shows the recommended elements and Figure \ref{fig:cs_results} shows the results.

\begin{table*}[ht]
\footnotesize
\centering
\caption{Recommended elements for the context used in the case study project.}
\label{tab:cs_rec_elements}
\begin{tabularx}{\linewidth}{@{}rX@{}}
\toprule
Element & Description \\ \midrule
test team & Small team with focused on development, not marketing \\
small team & Beta test group from community. Beta testers with access to bug database. Outsourcing test company to check devices from different platforms \\
general team details & Lead designer no present. Lack of artists. Experienced team who already worked together. No need to research, team have know-hall about tasks to complete. Full focused with no business meetings. \\
outsourcing & Outsourcing experienced people with technical expertise. Outsourcing PR / marketing. Outsourcing code and assets. \\
horizontal development & Follow a higher principle. Everyone may vet new ideas \\
development problems & Keep the pace and not ``crunch'' (work overtime) Lack of proper pipeline. Hard to solve things by yourself. Cutting or reworking features. \\
development process details & Lack of process or structure. Just code without planning. List of main features instead of a design document.  Share the development with the audience. No schedule, nor milestones, nor meetings nor design document nor technical plan Development flow \\
engine and tools & Use a tool for build distribution. Learn everything from scratch. \\
infrastructure & Self made engine and--or tools. Engine and--or legacy code limiting improvements. \\
project focus & Focus on the team strength. \\
scope & Changing scope. Developing self engine. \\
concept & Artists, level designers, programmers, and animators working together. Set of design principles. Brainstorming ideas based in a goal. Align thinking before prototyping phase. Research similar games. Stick with the game concept since the beginning. \\
brainstorming features & Brainstorming ideas finding ``fun factor''. \\
business tasks & Market study. Selling directly to customers. Pre-order program. Marketing strategy \\
design tasks & Procedurally generated instead of hand craft levels. AI specialists. \\
development iterations loop & Digital distribution allowing testers give feedback quickly and often. Heavy tested by non designers. Schedule iterations with buffers, that is, double or triple the time required for a task. Heavily focused on iterations and constant improvements in the game. Legacy problems being solved as the development goes on. \\
exploration phase & Prototyping a game without polishing, trying to find a good game play. Heavily prototyping and testing iterations. Avoid making experiments in production phase. \\
in-house tools development & Creating tools to aid developers with a new technology and avoid unknown bugs. Tools allowing non coders to change parameters and experiment with the game engine. The main goal is to create a new engine. Creating an IDE. Creating an Engine with experienced team. Creating tools to automatize tasks and processes. \\
milestones planning & Pre-order program to raise money allowing the team continue the development. Early access feedback. \\
multi-player construction & Specialist in multi-player development. \\
pitch & Pitch the game concept. \\
planning documentation & Production plan with buffers (more time) after each milestone. Cutting features, making less but better. \\
polish and refinements & Improving performance \\
professional feedback & Getting technical feedback. \\
prototyping & Simple prototyping to find ``fun factor''. Game being developed in parallel with the engine. \\
refactoring the development & Time expended with re-designs. Refactoring old code. \\
requirements and constraints & Game design documentation not updated. Define a list of features and tools. Port UI to other platform. \\
retrospective meeting & Research target audience. \\
testing & Beta test group. Multi-platform testing. Delivery builds quickly to test team. Long time beta testing the game. Get feedback from QA with volunteers. Feedback from testers not used properly. \\
users feedback & Working directly with gamers community using a on-line message board. Game shipped for one platform first. \\
vertical slice & Community working together creating assets. \\ \bottomrule
\end{tabularx}
\end{table*}

The \textbf{Trustworthiness} metric had 50\% of elements considered \textbf{similar} to what was used or applied in the development process. 28.13\% of the elements were not considered similar while 21.88\% were neutral.

\begin{quotation}
\noindent
\textbf{test team}
\textit{``We have a third party company who perform the tests before it reach the final client (...) the testers have access to the bug database.''}
\end{quotation}

\begin{quotation}
\noindent
\textbf{refactoring the development}
\noindent
\textit{``It happens often as the development adopt an evolutionary approach (...)''}
\end{quotation}

\begin{quotation}
\noindent
\textbf{retrospective meeting}
\noindent
\textit{``(...) retrospective meetings occur regularly with development team and the final client.''}
\end{quotation}

\begin{quotation}
\noindent
\textbf{general team details}
\noindent
\textit{``These (recommendations) do not applied in our context: The team had not the proper knowlegde in the beginning (...) nor had been worked together''}
\end{quotation}

The \textbf{Novelty} metric had 40.63\% of elements considered \textbf{new} by the development team. 34.38\% were considered already known and 25.00\% were neutral.

\begin{quotation}
\noindent
\textbf{in-house tools development}
\noindent
\textit{``Although these practices were not used (by us) today (...) we found it a good surprise.''}
\end{quotation}

The \textbf{Serendipity} metric had 37.50\% of elements considered \textbf{surprisingly good} for the project. 25\% were considered not good and 37.50\% were neutral.

\begin{quotation}
\noindent
\textbf{engine and tools}
\noindent
\textit{``We do not have a build tool (...) most of the techniques applied are learned from scratch.''}
\end{quotation}

The \textbf{Utility} metric had 59.38\% of elements considered \textbf{useful} to be used in the actual process. 12.50\% were not considered good and 28.13\% were neutral.

The \textbf{Risk} metric had 28.13\% considered \textbf{risky} if added to the project. 46.88\% were considered not risky and 25\% were neutral. However, among these elements, none received a ``5'' as a answer.

The team cited a interesting point, as the project had already more than three years of development, many of the activities used were the results of ``trials and errors''. Using the recommendation system, some of these errors could have been avoided :

\begin{quotation}
\noindent
\textit{``This practices could be used since the beginning, for example, prototyping, test team, bug control, small team, building tools and do less but better.''}
\end{quotation}

Generally speaking, the team emphasized the the following subset of activities does not applied to their project: user community, gathering resources, marketing, fun factor related practices and any activity multi-player. From the recommendation system, there were two details to consider. Firstly, some elements had details from more than one project, which were merged together. Therefore, a recommended element may contradict itself, showing a activity that makes another impossible. Secondly, the context has some particular elements not contained in the analyzed postmortems. For this reason, a few elements were not considered similar nor useful.


\section{Discussion}
\label{sec:discussion}

The use of postmortems to analyze video game projects is not new but used with another goal it this article: to extract the processes used by developers. Even though the intent of postmortems is not to document processes, it is still possible to extract data relevant to the processes and, with many postmortems, assemble activities to create a flow of activities realized by the team.

For this article, we extracted 55 processes for postmortem analysis. These processes document what occurred during the development of the games. Thus, the extracted processes, as well as the extraction system, may serve as a tool to analyze past projects. For example, a developer, who had developed 10 games and wants to make a retrospective of all projects, can use our system to extract 10 processes and generate their graphical representations and make comparisons, group similar projects, and even define metrics and evaluate the processes, projects, and teams.

Something that may help to solve some issues regarding the postmortem analysis and help the developers community, would be a creation of an \textbf{experiences repository} where finalized projects reports would be stored using a more rigid structure, describing errors and hits during the project execution.


We did not find a source related to a definition of video game project context. We created a way to define video game project contexts, based on literature, books, and postmortems. We defined 61 video game variables, divided into 6 categories. This set of variables is not immutable but is a basis for future extensions. Moreover, we encourage aggregation or refactoring variables because such operations do not interfere with how the recommending system works.

The recommendation system is the main contribution of this article. The aim of the generated processes is to help the development team in planning a game project, even before the preproduction phase. The resulting processes help developers to make decisions regarding the development phases or serving as a high level process to be followed.

For example, before the development starts, with the game context defined, the context form is filled by game developers. The recommendation system creates a new process using similar projects as a basis. The development team or responsible developer may analyze all the problems that occurred in other, similar games. With this data, it is possible to draw plans that prevent such problems and apply activities that went right.

The system evaluation, although with relatively low measures, results in good acceptance by the game developers. Developers who wrote postmortems acknowledged the similarity of the extracted and recommended processes. The developers related that many of the activities presented by the recommendation system were already part of the processes used by their teams. It emphasizes that the premise of this article, ``learning with the past'', can be a way to avoid video game development problems.

\section{Threats to Validity} \label{sec:threats}

There are concerns about the generated processes, given that the postmortem main purpose is not describe processes' elements. It took us to carry out the validation of both, the extracted and recommended processes.

As pointed in a feedback, the recommended process cannot define specifics tasks to be followed by developers. Again, it is because the information contained in postmortems does not have such specificity. Nonetheless, we argue that it is the best source of information about development experiences.

Most of the work of creating a process visualization had to have be made manually, which cost much time. Although the process visualization is built mainly by scripts, its flow is entirely made by hand. It happened because we did not stored the meta-data about elements' flow, which can be done in future extensions.

We performed tests using a dataset with 55 samples against 61 contextual properties. This might cause noise in the dataset, that is, \textit{Overfitting} issues \cite{robillard2014recommendation}.


\section{Related Work} \label{sec:related}

Postmortems was used by researchers as a primary source of information about game development. Petrillo \textit{et al.} \cite{Petrillo:2008, Petrillo:2009} extracted, with postmortem analysis, the most common problems faced by developers in game industry. They argued that these problems occur in the traditional software industry as well. Still, the authors stated that game developers suffer mainly from managerial problems, not technical ones.

In a complementary research, Petrillo \textit{et al.} list the most common practices adopted in game projects. The results showed that, even if informally, game developers are adopting agile practices and, for this reason, implanting agile methods may occur naturally \cite{Petrillo:2010}.

Another research, done by Washburn \textit{et al.} \cite{Washburn:2016}, regarding development practices and postmortems, analyzed  155 articles extracting the characteristics and pitfalls from the game development. The result was a set of good practices for game developers.

Murphy-Hill \textit{et al.} \cite{Murphy-Hill:2014} realized 14 qualitative interviews and 364 quantitative, with professionals from game development area. The authors highlighted the dubious agile methods adoption, that is, the use of the word ``agile'' to justify the lack of process. Moreover, Pascarella \textit{et al.} \cite{Pascarella2018} extended this work assessing how developing games is different from developing non-game systems by performing analysis in both Open Sourced repositories types. They found that developing video games differs from developing traditional software, in many aspects, like:
the way resources evolve during the project;
the expertise required by the developers;
the way the software is tested;
how bugs are fixed;
how release planning is followed and
regarding requirements handling.

O'Hagan \textit{et al.} did a literature review of 404 articles, from industry and academy, showing a total of 356 software processes, grouped in 23 models, of which 47\% were agile and 53\% hybrid \cite{OHagan:2014}. Still, O'Hagan \textit{et al.} made a case study in game industry exploring the role of software process in game development. The results showed that there is no good practices model for game development and suggested to create such model, based on ISO/IEC 29110 \cite{OHagan:2015}.

Kanode \textit{et al.} also explored the role of software engineering in game development. For the authors, game development has unique characteristics and problems, and software engineering practices may help to overcome these difficulties. For example, implanting agile methods in prep-production phase and managing the large amount of game assets \cite{Kanode:2009}.

Regarding Recommendation Systems, there are several attempts to provide better tools to developers. For example, recommendation of developments artifacts~\cite{Cubranic:2003, Holmes:2008, Holmes:2005, Zimmermann:2004}, contextual information~\cite{Ponzanelli:2017}, libraries~\cite{ouni2017search}, methods~\cite{velasquez2017kontun}, commands~\cite{gasparic2017graphical} and even pull-requests commenter \cite{jiang2017should}.

These related works showed the game industry problems, suggested the use of good practices and provide recommendation of different development artifacts. However, there is no mention, as far as we know, about a implementation and validation of a system to aid developers, recommending a set of activities and practices in the form of a process, based on similar projects context.

\section{Conclusions}
\label{sec:conclusion}

Building a video game demands a large number of activities and choosing these activities is arduous and demands experience.
This article presented an approach and a recommendation system, based on video game development postmortems, to suggest software processes using contexts and similarity degrees. The recommendation system helps video game developers to conduct new projects by suggesting a set of activities in a form of a process. It learns from previous game development experiences within similar contexts. Moreover, provides to video game developers an approach to reflect about the processes used in game industry.

This article thus presents three main contributions.
It first describes the creation of a database of game development processes from the analysis of 55 postmortems.
The second identifies video game project characteristics, like team attributes and technical details, to model projects contexts that work as input for the recommendation system.
The third and final one describes and validates a recommendation system capable of generating processes based on previous projects with similar contexts.

The quantitative results of the validation of the recommendation system showed an average \textit{precision} of 31.58\%; good \textit{accuracy} of 67.59\%; and excellent \textit{specificity} of 92.48\%. However, other metrics had lower outcomes, like \textit{recall} at only 8.11\%.
These results showed that a better technique must be provided, in order to filter the recommended processes' elements: provide more elements with more relevance. The former is achievable by adding a larger numbers of postmortems in the database while the latter requires using better classification algorithms, improving the video game context categories, and enriching the elements of the processes.

The qualitative validation provided by video game developers showed that our approach does extract similar process from postmortems as well our recommendation system suggests useful new processes.

Finally, through a case study with a team of developers, we showed that video game development teams may, transparently, improve their software processes during the course of game development. Thus, our recommendation system can recommend video game development processes that help game developers to avoid possible problems.

As future work, we intend to improve our recommendation system by adding more postmortems and analyzing their activities. We will gather more resources and feedbacks from video game developers to improve our model. Moreover, we will improve the method used to generate processes visualization as well as the user interface for the recommendation system. Further, we will add more meta-data in process' elements, like order and flow. With this we hope cut most of the hand working and provide a completely automated processes recommendation system.

\section*{Acknowledgements}

We give our thanks to the authors of the postmortems for sharing their cases as well for their feedbacks. This work has been supported by the National Council for Scientific and Technological Development of Brazil.

\section*{References}

\bibliographystyle{elsarticle-num}
\bibliography{ist-main}

\end{document}